\documentclass[useAMS,usenatbib,usegraphicx]{mn2e}
\usepackage{xspace,amsmath}
\usepackage{color,epsfig,amssymb,lscape}
\usepackage{util}
\usepackage{array,colortbl}
\usepackage[colorlinks=true,linkcolor=magenta,citecolor=blue,breaklinks=true]{hyperref}
\usepackage{ifpdf}
\usepackage{url}

\ifpdf
\usepackage{epstopdf}
\else
\usepackage[dvips]{graphicx}
\usepackage{lscape}
\fi

 \newcommand{\hii}{\relax \ifmmode {\mbox H\,{\scshape ii}}\else H\,{\scshape ii}\fi}
\newcommand{\mi}{\relax \ifmmode {\mu{\mbox m}}\else $\mu$m\fi}
\newcommand{\ha}{\relax \ifmmode {\mbox H}\alpha\else H$\alpha$\fi}
\newcommand{\hb}{\relax \ifmmode {\mbox H}\beta\else H$\beta$\fi}

\newcommand{\sii}{\relax \ifmmode {\mbox S\,{\scshape ii}}\else S\,{\scshape ii}\fi}
\newcommand{\siii}{\relax \ifmmode {\mbox S\,{\scshape iii}}\else S\,{\scshape iii}\fi}
\newcommand{\nii}{\relax \ifmmode {\mbox N\,{\scshape ii}}\else N\,{\scshape ii}\fi}
\newcommand{\oi}{\relax \ifmmode {\mbox O\,{\scshape i}}\else O\,{\scshape i}\fi}
\newcommand{\oii}{\relax \ifmmode {\mbox O\,{\scshape ii}}\else O\,{\scshape ii}\fi}
\newcommand{\oiii}{\relax \ifmmode {\mbox O\,{\scshape iii}}\else O\,{\scshape iii}\fi}
\newcommand{\neiii}{\relax \ifmmode {\mbox Ne\,{\scshape iii}}\else Ne\,{\scshape iii}\fi}

\newcommand{\rdostres}{\relax \ifmmode {\,\mbox{R}}_{\rm 23}\else \,\mbox{R}$_{\rm 23}$\fi} 

\newcommand{\gsim}{\hbox{\rlap{\lower.55ex\hbox{$\sim$}} \kern-.3em
\raise.4ex \hbox{$>$}}}
\newcommand{\lsim}{\hbox{\rlap{\lower.55ex\hbox{$\sim$}} \kern-.3em
\raise.4ex \hbox{$<$}}}

\title[Model-based T$_e$-consistent abundances]{Deriving model-based T$_e$-consistent chemical abundances in ionised gaseous nebulae}

\author[E. P{\'e}rez-Montero]
       {E. P{\'e}rez-Montero\thanks{E-mail:epm@iaa.es}\\
Instituto de Astrof\'\i sica de Andaluc\'\i a. CSIC. Apartado de correos 3004. 18080, Granada, Spain.}
        
\date{2014}

\pagerange{\pageref{firstpage}--\pageref{lastpage}}
\pubyear{2013}

\begin{document}

\maketitle

\label{firstpage}

\begin{abstract}
The derivation of abundances in gaseous nebulae ionised by massive stars
using optical collisionally excited emission lines is studied in this work comparing 
the direct or $T_e$ method with updated grids of photoionisation models
covering a wide range of input conditions of O/H and N/O abundances and ionisation
parameter.
The abundances in a large sample of compiled objects with at least one auroral line 
are re-derived and later compared with the $\chi^2$ weighted-mean 
abundances from the models. The agreement between the abundances
using the two methods both for O/H and N/O is excellent with no additional
assumptions about the geometry or physics governing the \hii\ regions.
Although very inaccurate  model-based O/H are obtained
when no auroral lines are considered, this can
be overcome assuming empirical laws between O/H, log $U$, and N/O
to constrain the considered models.
In this way, for 12+log(O/H) $>$ 8.0, a precision better than 0.1dex
consistent with the direct method is attained. For very low-$Z$, models
give higher O/H values and a high dispersion, possibly owing
to the contamination of the low-excitation emission-lines. However, in this
regime, the auroral lines are usually well-detected.
The use of this procedure, in a publicly available script, {\sc hii-chi-mistry},
leads to the derivation of abundances in faint/high redshift objects consistent
with the direct method based on CELs.

\end{abstract}


\section{Introduction}

Metalllicity ($Z$) is one of the most relevant quantities to
correctly characterize the nature of astrophysical objects, including
from planets and stars up to galaxies. For instance, in the early Universe an accurate
determination of $Z$ as a function of the cosmic age and other
properties of galaxies renders very valuable
information about the evolution of the Universe itself. At these
distances, the collisionally excited lines (CELs) emitted by different
chemical species in the gas ionised by massive episodes of 
star-formation is almost the unique way to access to this information. 
Owing to the strong dependence of CELs on electron temperature ($T_e$), the method
used to derive chemical abundances using these lines relies necessarily on the previous 
determination of $T_e$, using the so-called direct method.
In the case of the optical spectral range, the temperature can be estimated
using specific auroral-to-nebular emission line ratios, such
as [\oiii] I(5007\AA)/I(4363\AA). 

However, as auroral emission lines are much fainter than nebular
strong lines, a direct estimation of $T_e$ is
difficult in faint/distant objects or in the high-$Z$ regime, where
the cooling is more efficient and the temperature is then lower.
In this case, strong-line methods based only on the nebular CELs are used.
In order to provide a $Z$ absolute scale consistent with the direct
method, some authors have provided empirical calibration of these strong-line
methods (e.g. \citealt{Pilyugin05, PP04, PMD05}). This methodology
has the advantage that is well correlated with direct observations but, on the
contrary, can lead to biases in the scale towards the objects with
a good measurement of the auroral lines \citep{stasinska10}. In this case,
a calibration of the strong-line methods based on models ensure that all conditions
can be envisaged. The main problem is that many of the
most widely used calibrations based
on models (e.g. \citealt{McG91, CL01, KD02}) give systematic
overabundances in relation with the direct method. These
differences can be as high as 0.7dex depending
on the models and the $Z$ regime (see for a discussion on this \citealt{KE08}).

In the same sense, the direct method based on CELs has been
widely questioned based on the controversial results
coming from other observational techniques in the same objects
where the CELs have been detected. 
For instance, the use of recombination lines (RLs) emitted by heavy elements, 
which are around 10$^4$ times fainter than CELs, but they do not depend on temperature, 
lead to abundances 0.1-0.3dex higher both in Giant \hii\ regions (e.g. \citealt{esteban09} 
and references therein) and even higher in planetary nebulae (e.g. \citealt{liu10} and
references therein). This discrepancy is not probably due to the use of old
sets of atomic data for the RLs \citep {fang13}. 
On the other hand, the abundances derived from OB supergiants
give controversial results about the value of the discrepancy with abundances
from CELs. Some works point to little difference with CELs (e.g. NGC300, \citealt{bresolin09}),
and while some others point to values quite similar to those obtained from RLs (e.g. \citealt{SDS11,FP12}).

We can find different explanations in the literature to account
for these differences, including the presence of fluctuations of 
temperature \citep{peimbert67}, which makes the
integrated $T_e$ derived from CELs ratios to be systematically
overestimated and, hence, lead to an underestimation of the
metallicity. Other possible cause is found in the existence of chemical
inhomogeneities \citep{tsamis05, dors13}  or, more recently, in a 
possible asymmetric energy distribution
for the free electrons in the ionised gas (the $k$ distribution, \citealt{kappa}).

The existence of these discrepancies has been used as an argument
to justify the disagreement found between the calibrations of
strong-line methods based on models and based on observations of CELs. 
Nevertheless, the physical causes thought
to be responsible for these discrepancies are not included in the models and
even some tailor-made models for low-$Z$ objects, in which [\oiii]
4363 \AA\ is visible give chemical abundances in good agreement with those
derived from the direct method (e.g. \citealt{PMD07,PM10, Dors11}).

This work studies if photoionisation models can lead
to a determination of chemical abundances consistent with the
calculations obtained using the direct method both with and
without measuring auroral emission lines. To this aim,
Section 2 presents a sample of ionised gaseous nebulae with
good measurements of their CELs, including at least one auroral emission line,
and their O/H and N/O are re-calculated using expressions consistent with
the code {\sc pyneb} and updated sets of atomic data. Section 3 describes a large grid of
photoionisation models covering a wide range of properties in O/H, N/O,
and log $U$. and a procedure based on a $\chi^2$-weighted
mean is presented. Section 4 analyses the consistency of the
abundances obtained using the two methods, and discusses what
happens when the number of available emission lines is limited in the models.
Finally, Section 5 summarises the results and presents the conclusions.


\section{Data description}

This study uses the compilation of emission-line fluxes in ionised gaseous nebulae 
carried out by \cite{RAM13} to provide a data sample
that can be later compared with the results obtained from models. 
This sample comprises 550 objects collected from the literature with
the measurement with good S/N of at least one auroral emission-line and 
hence an empirical estimate of $T_e$. 
It is an heterogeneous sample of pure star-forming 
low-density (i.e. $n_e <$ 1000 cm$^{-3}$) emission-line objects in the
Local Universe including \hii\ galaxies,
giant extragalactic \hii\ regions, and diffuse \hii\ regions in our own Galaxy and
in the Magellanic Clouds. 
This sample is the largest available so far with $T_e$ and covers 
a wide range in O/H from 12+log(O/H) = 7.1
(SBS0335-052) up to 9.1 (H13 in NGC628)] and N/O [from log(N/O) = -1.95 (NGC5253)
up to -0.14 (NGC5236)].
The emission-lines measured by means
of integrated spectroscopy of the brightest \hii\ regions were
consistently reddening corrected by \cite{RAM13} and later used
to calculate chemical abundances according with the direct method.

In this sample, an empirical estimate of the [\oiii] $T_e$
by means of the I(5007)/I(4363) can be performed in 497 objects.
In 76 objects, 17 of them also belong to the sample with
t[\oiii], the [\nii]$T_e$ can be derived empirically using
the I(6584)/I(5755). The leftover 37 objects have an estimate of 
$T_e$ of [\siii] by means of the ratio I(9532)/I(6312).
Both t[\nii] and t[\siii] can be derived more easily than t[\oiii] in high-$Z$
objects as the emissivity of the corresponding emission-lines depends
less on $T_e$ \citep{bresolin07}.

\subsection{Derivation of physical properties and chemical abundances}

For this work, all $n_e$, $T_e$ and ionic abundances of O$^+$,
O$^{2+}$ and N$^+$ were re-calculated using expressions derived
using non-linear fittings to the results obtained from
the emission-line analysis software {\sc pyneb} v0.9.3 \citep{pyneb} 
covering the conditions of the studied sample
as described below and
with the most updated sets of atomic coefficients in agreement with
the photoionisation models. 
The expressions were obtained using arbitrary sets of 
input emission-line intensities covering the conditions of the data.
These formulae are provided to ease the reproducibility
of the calculations, the error analysis and their applicability for large
data samples using different software.

Electron densities are necessary for the derivation of chemical abundances
of ions of the type np$^2$, such as O$^+$. These densities were derived using
the following emission-line ratio:

\begin{equation}
R_{S2} = \frac{I(6716)}{I(6731)}
\end{equation}

\noindent with all wavelengths in this and hereafter expressions 
are given in Angstr\"oms.  As in \cite{H08} the following expression
is proposed to derive the electron density:

\begin{equation}
\textrm{n$_e$([\sii])} = 10^3 \cdot \frac{R_{S2} \cdot a_0(t) + a_1(t)}{R_{S2} \cdot b_0(t) + b_1(t)}
\end{equation}

\noindent with $n_e$ in units of cm$^{-3}$ and $t$ in units of
10$^4$ K. The $T_e$ used here is that of t(N$^+$)
calculated as described below. Using the appropriate fittings and {\sc pyneb} with 
collision strengths from \cite{TZ10} gives these polynomial fittings 
to the coefficients

\[ a_0(t) = 16.054 - 7.79/t - 11.32\cdot t \]
\[ a_1(t) = -22.66 + 11.08/t + 16.02\cdot t \]
\[ b_0(t) = -21.61 + 11.89/t + 14.59\cdot t \]
\begin{equation}
b_1(t) = 9.17 - 5.09/t - 6.18\cdot t 
\end{equation}

This expression fits the density calculated by {\sc pyneb} better
than a 1\% for temperatures in
the range 0.6 $< t_e <$  2.2 and densities in the range 10 $< n_e <$ 1000.
 
The $T_e$ of [\oiii] was calculated from the emission-line
ratio: 

\begin{equation}
R_{O3} = \frac{I(4959)+I(5007)}{I(4363)}
\end{equation}

Using {\sc pyneb} the following
non-linear fitting for $n_e$ = 100 cm$^{-3}$:

\begin{equation}
\textrm{t([\oiii])} = 0.7840 - 0.0001357\cdot R_{O3} + \frac{48.44}{R_{O3}} 
\end{equation}

\noindent in units of 10$^4$, valid in the range $t$ = 0.7 - 2.5
and using collisional strengths from \cite{ak99}. This fit gives
precisions better than 1\% for 1.0$<$ t([\oiii]) $<$ 2.5, and
better than 3\% for 0.7 $<$ t([\oiii]) $<$ 1.0. It was calculated
for a density of 100 cm$^{-3}$, but considering a density
of 1000 cm$^{-3}$ reduces the temperature only in a 0.1\%.

The$T_e$ of [\nii] was calculated using the
ratio:

\begin{equation}
R_{N2} = \frac{I(6548)+I(6584)}{I(5755)}
\end{equation}

\noindent that, with the corresponding fitting leads to
the expression:

\begin{equation}
\textrm{t([\nii])} = 0.6153 - 0.0001529\cdot R_{N2} + \frac{35.3641}{R_{N2}} 
\end{equation}

\noindent also in units of 10$^4$ K, in the range $t$ = 0.6 - 2.2
using collision strengths from \cite{T11}. This fit gives a precision better
than 1\% in the range 0.7 $<$ t([\nii]) $<$ 2.2 and better than 3\% in
the range 0.6 $<$ t([\nii]) $<$ 0.7. It was calculated for a density
of 100 cm$^{-3}$, but for a density of 1000 cm$^{-3}$, the temperature
is reduced in less than a 1\%.

The calculation of chemical abundances from collisionally excited lines
(CELs) depends strongly on the adopted $T_e$.
Therefore, it is fundamental to properly assign the temperature in the zone where 
each ion is. For this work the same criterion is adopted as \cite{garnett92} and it
is considered that the ion temperature can be taken as the corresponding line
temperature, so t(O$^{2+}$) $\approx$ t([OIII]) and 
t(N$^+$) $\approx$ t([\nii]). In those objects  
without a direct estimation of t(N$^+$) it is used the
following expression derived from the same photoionisation models described in
the next section:

\begin{equation}
\textrm{t(N}^+) = \frac{1.452}{1/\textrm{t(O}^{2+})+0.479}
\end{equation}

While the temperature of O$^+$ was calculated using the 
following expression from the same set of models,
valid for all electron densities lower than the critical value:

\begin{equation}
\textrm{t(O}^+) = \frac{1.397}{1/\textrm{t(O}^{2+})+0.385}
\end{equation}

In the case of the 37 objects whose unique auroral line is [\siii] 6312 \AA\ 
it can be used the following emission-line ratio:

\begin{equation}
R_{S3} = \frac{I(9069)+I(9532)}{I(6312)}
\end{equation}

\noindent what leads to the following fitting in the range
$t$ = 0.6 - 2.5 using the collision strengths from \cite{HRS12}

\begin{equation}
\textrm{t([\siii])} = 0.5147 + 0.0003187\cdot R_{S3} + \frac{23.64041}{R_{S3}} 
\end{equation}

\noindent with a precision better than 1\% in the range
0.6 $<$ t([\siii]) $<$ 1.5, and better than 3\% up to values
t([\siii]) = 2.5. These values enhance in less than a 3\% when the
considered density goes from 100 to 1000 cm$^{-3}$.
Then, assuming that t(S$^{2+}$) $\approx$ t([\siii]) and
considering the results from models:

\begin{equation}
\textrm{t(O}^{2+}) = 1.0807\cdot \textrm{t(S}^{2+}) - 0.0846
\end{equation}

The chemical abundance of O$^+$ was derived in all
objects with the relative intensity of [\oii] 3726, 3729 \AA\
emission lines to H$\beta$ and the corresponding temperature using
the following expression obtained from fittings to {\sc pyneb} using
the default collision strengths from \cite{pradhan06} and \cite{tayal07}:

\[
12+\log\left(\frac{O^+}{H^+}\right) = \log\left(\frac{I(3726)+I(3729)}{I(H\beta)}\right) +
\]
\begin{equation}
+ 5.887 + \frac{1.641}{t(O^{+})} - 0.543\cdot\log(t(O^{+})) + 0.000114\cdot n_e
\end{equation}

\noindent with a precision better than 0.01dex in the temperature
range 0.7 $<$ t(O$^+$) $<$ 2.5 and density of 100 cm$^{-3}$. For
a density of 1000 cm$^{-3}$ the precision is better than 0.02dex.
Regarding O$^{2+}$, its chemical abundance was derived 
using the relative intensity of [\oiii] 4959, 5007 \AA\
emission lines to H$\beta$ and the corresponding temperature using
the following expression obtained from fittings to {\sc pyneb}:

\[
12+\log\left(\frac{O^{2+}}{H^+}\right) = \log\left(\frac{I(4959)+I(5007)}{I(H\beta)}\right) +
\]
\begin{equation}
+ 6.1868 + \frac{1.2491}{t(O^{2+})} - 0.5816\cdot\log(t(O^{2+}))
\end{equation}

\noindent with a precision better than 0.01dex in the temperature range
0.7 $<$ t(O$^{2+}$) $<$ 2.5. A change in the density from 10 to 1000 cm$^{-3}$
implies a decrease of less than 0.01dex in the derived abundance.

Assuming that all the oxygen is in the two above mentioned states of
ionisation, the total abundance of oxygen can be calculated adding these
two abundances. In the case of nitrogen, N$^+$ abundance
can be estimated using relative intensity of [\nii] 6548, 6584 \AA\AA\ 
to H$\beta$ with its corresponding temperature and this expression:

\[
12+\log\left(\frac{N^+}{H^+}\right) = \log\left(\frac{I(6548)+I(6584)}{I(H\beta)}\right) +
\]
\begin{equation}
+ 6.291 + \frac{0.90221}{t(N^{+})} - 0.5511\cdot\log(t(N^{+})) 
\end{equation}

\noindent with a precision better than 0.01dex in the temperature
range 0.6 $<$ t(N$^+$) $<$ 2.2. It decreases less than 0.01dex when
the considered density goes from 100 to 1000 cm$^{-3}$.

The N/O ratio was then derived assuming the approximation:

\begin{equation}
\frac{N^+}{O^+} \approx \frac{N}{O}
\end{equation}

\section{Model-based abundance derivation}

\subsection{Description of the models}

A grid of photoionisation models was performed to provide a complete
set of emission-line intensities as a function of O/H and N/O
at different assumed conditions of excitation. This work uses
the synthesis spectral code {\sc cloudy} v13.03 \citep{cloudy}, which calculates the
emergent spectrum from a one-dimensional distribution of gas and dust
irradiated with an arbitrary input spectral energy distribution (SED).
This study utilises {\sc popstar} \citep{popstar} synthesis evolutionary models as cluster ionising SED 
assuming an instantaneous burst with an age of 1 Myr with an initial mass function
of \cite{chabrier} and using in each model the metallicity assumed for the gas,
as scaled to the solar value with the oxygen abundance.
The models assume a distance between the ionising source and the
inner face of the gas at which the
geometry is plane-parallel with a constant electron density of 100 cm$^3$.
The calculation is stopped when the ratio of ionised hydrogen atoms is less than 98\%.
Possible excitation differences owing to varying age, mass, or geometrical conditions in a wide range of
possible scenarios were covered using variations  
of the ionisation parameter, which can be defined as:

\begin{equation}
\log U = \frac{Q(H)}{4\pi r^2 n c}
\end{equation}

\noindent where $Q(H)$ is the number of ionising photons in s$^{-1}$, $r$ is the outer 
radius of the gas distribution in cm, $n$ is the density of particles in cm$^{-3}$,
and $c$ is the speed of the light in cm$\cdot$s$^{-1}$. 
The grid considers values of log $U$ from -1.50 until -4.00 in steps of 0.25dex.

The models consider default grain properties and relative abundances
(i.e. using a \cite{mathis77} size distribution and a dust-to-gas mass
ratio of 7.5$\cdot$10$^{-3}$).
The chemical composition of the gas is traced with
the total oxygen abundance, for which 21 different values
for which the models take values in the range 12+log(O/H) = [7.1,9.1] in
steps of 0.1dex. The rest of elements were scaled following the
solar proportions given by \cite{asplund} and considering 
the {\sc Cloudy} default
depletion factors. Only in the case of nitrogen, the models
consider variations of the N/O ratio to take the dependence
of the [\nii] optical emission-lines on N/H abundance into account. 
This grid assumes  17 different
values of the ratio log(N/O) in the range [0.0,-2.0] in steps
of 0.125dex. Then the total number of models in this grid is 
11 $\times$ 21 $\times$ 17 = 3927 \footnote{The models used in this paper are 
stored on the 3MdB database (ref. HII\_CHIm) (\citealt{3mdb}, Morisset in prep.). More information on the 
3MdB project and the models can be found in \url{https://sites.google.com/site/mexicanmillionmodels/}}.

\begin{figure*}
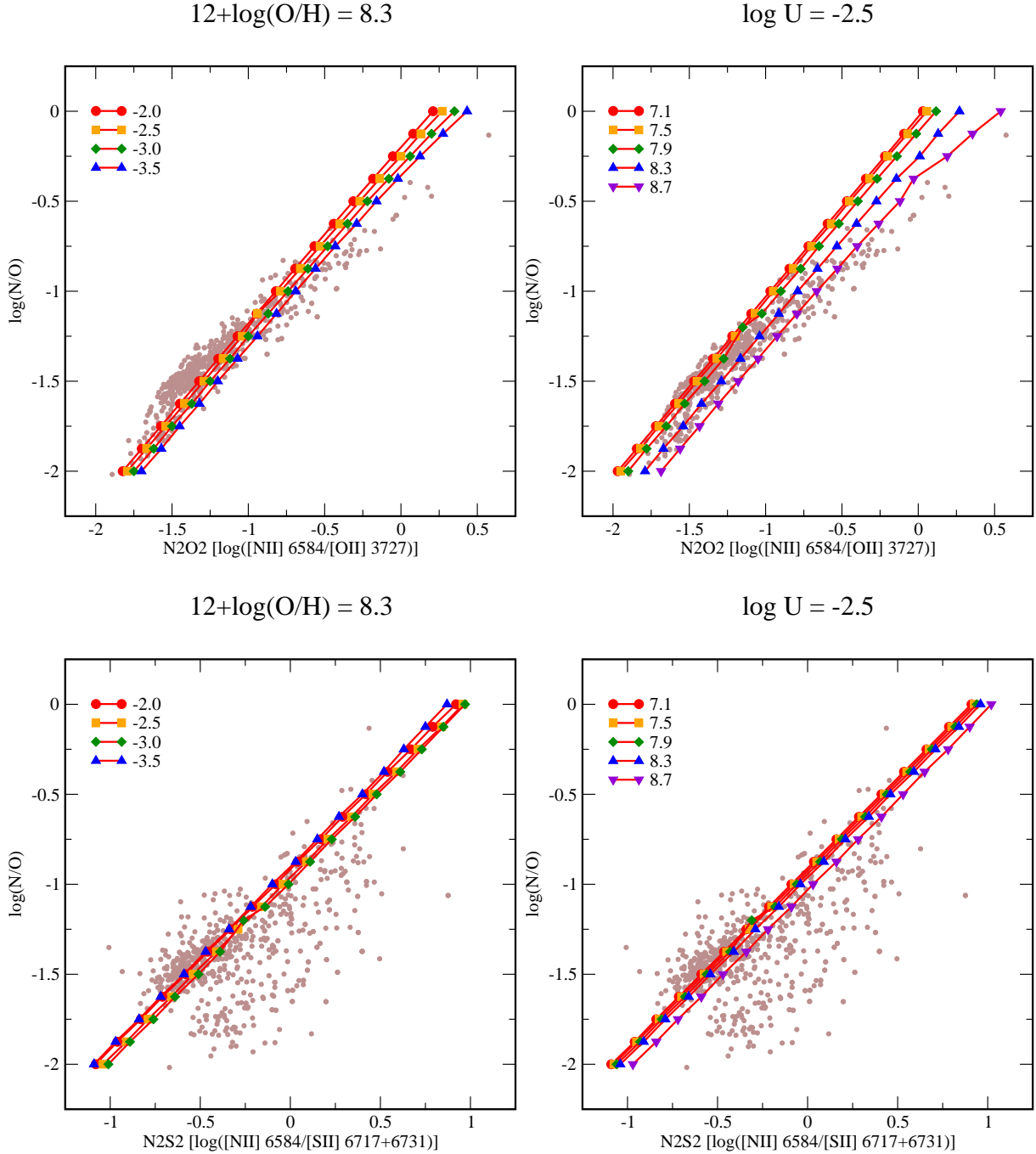


\begin{minipage}{180mm}

\centerline{
\psfig{figure=N2O2_NO_OH8.3.eps,width=8cm,clip=}
\psfig{figure=N2O2_NO_U2.5.eps,width=8cm,clip=}}
\vspace{0.5cm}
\centerline{
\psfig{figure=N2S2_NO_OH8.3.eps,width=8cm,clip=}
\psfig{figure=N2S2_NO_U2.5.eps,width=8cm,clip=}}

\caption[]{Relation between N/O and the N2O2 parameter in the upper row and the N2S2 parameter
in the lower row for the sample of objects described in this work. Plots in the left column also show the
results from models for different values of log $U$ at a fixed 12+log(O/H) = 8.3. In the right column,
the models have different values of metallicity at a fixed log $U$ = -2.5.}
\label{N2O2S2}

\end{minipage}
\end{figure*}

\subsection{Derivation of model-based properties}

The grid of models described in the above subsection can be used
to derive O/H, N/O, and log $U$ from the available optical CELs using
a $\chi^2$-based methodology. 
The calculation of the model-based chemical abundances
and ionisation parameter is based on the 
the predicted extinction-corrected intensities
relative to H$\beta$ of the emission lines [\oii] 3727 \AA,
[\oiii] 4363, 5007 \AA, [\nii] 6584 \AA, and
[\sii] 6716+6731 \AA\ . Notice that neither [\oiii] 4959 \AA\
nor  [\nii] 6548 \AA\ are included because the fluxes of these lines are
in a fixed relation with other emission lines already considered
[e.g. I(5007)/I(4959)=2.98, I(6584)/I(6548) = 3.05 \citep{sz2000}]

In a first step, for a given
set of observed extinction-corrected emission line intensities,
these are compared with the emission-lines from the models in order to estimate N/O.
According to \cite{PMC09}, some emission-line ratios between [\nii] and
other low-excitation emission-line, such as [\oii] (N2O2) or [\sii] 
(N2S2) depend basically on N/O. The apparent dependence
of these calibrators on metallicity is mainly due to the O/H vs. N/O relation in
the regime of production of secondary nitrogen.
However, this relation presents
a huge dispersion as N/O can also depend on star-formation history \citep{molla06}
and its relation with O/H be altered with the action of hydrodynamical processes such as the inflow of
metal-poor gas \citep{edmunds90, KH2005}. These ratios also have the
advantage that they do not depend on the excitation of the gas
\citep{KD02}.
In figure \ref{N2O2S2}, it is shown the relation between the two considered ratios
(N2O2 = log([\nii]/[\oii]); N2S2 = log([\nii]/[\sii]) and N/O for the sample of objects
described in section 2 and showing the results of some models to illustrate the dependence of these
ratios with O/H and with log U.

The final N/O value is then calculated as the weighted sum of the
N/O values in each model using the following expression:

\begin{equation}
\log(N/O)_f = \frac{\sum_i \log(N/O)_i/\chi_i}{\sum_i 1/ \chi_i}
\end{equation}

\noindent where $i$ are the different considered models from the grid, 
log(N/O)$_i$ are the values of log(N/O) in each model and $\chi_i$ are calculated as:
\begin{equation}
\chi_i^2 =\sum_j \frac{(O_j - T_{ji})^2}{O_j}
\label{chi}
\end{equation}

\noindent being $O_j$ and $T_{ji}$ the observed and model-based values, respectively, for the 
considered emission-line dependent ratios. In this case, N2O2, N2S2 and R$_{O3}$. 
R$_{O3}$ is also used since 
the N/O ratio also depends on $T_e$.

An error can also be derived using the following expression:
\begin{equation}
(\Delta\log(N/O))^2 = \frac{\sum_i \log((N/O)_f - \log(N/O)_i)^2/\chi_i}{\sum_i 1/\chi_i}
\label{error}
\end{equation}

In a second step, once N/O is estimated the grid is limited to those models
with the closest N/O values to the N/O$_f$ previously estimated 
(at this point, the number of models is reduced to 11$\times$21$\times$2 = 462)
and a new iteration is
made in order to derive the final values for O/H and log$U$ using a similar
expression to that described above:

\begin{equation}
12+\log(O/H)_f = \frac{\sum_k (12+\log(O/H))_k/\chi_k}{\sum_k 1/\chi_k}
\end{equation}
\begin{equation}
\log U_f = \frac{\sum_k \log U_k/\chi_k}{\sum_k 1/\chi_k}
\end{equation}

\noindent with $k$ $<$ $i$, as the number of models was limited to
those approaching the most to the derived N/O, and the values for $\chi$
are determined using the same expression as equation \ref{chi} but
using as observables the ratios R$_{O3}$, [\oii]/\hb, [\oiii] 5007/\hb,
[\nii]/\hb, and [\sii]/\hb. The errors for the final values
are also estimated using expressions similar to equation \ref{error}.

This procedure has been programmed in {\em python} language in a 
publicly available script called {\sc hii-chi-mistry}\footnote{In the web page
\url{http://www.iaa.es/~epm/HII-CHI-mistry.html}}.

\begin{figure*}
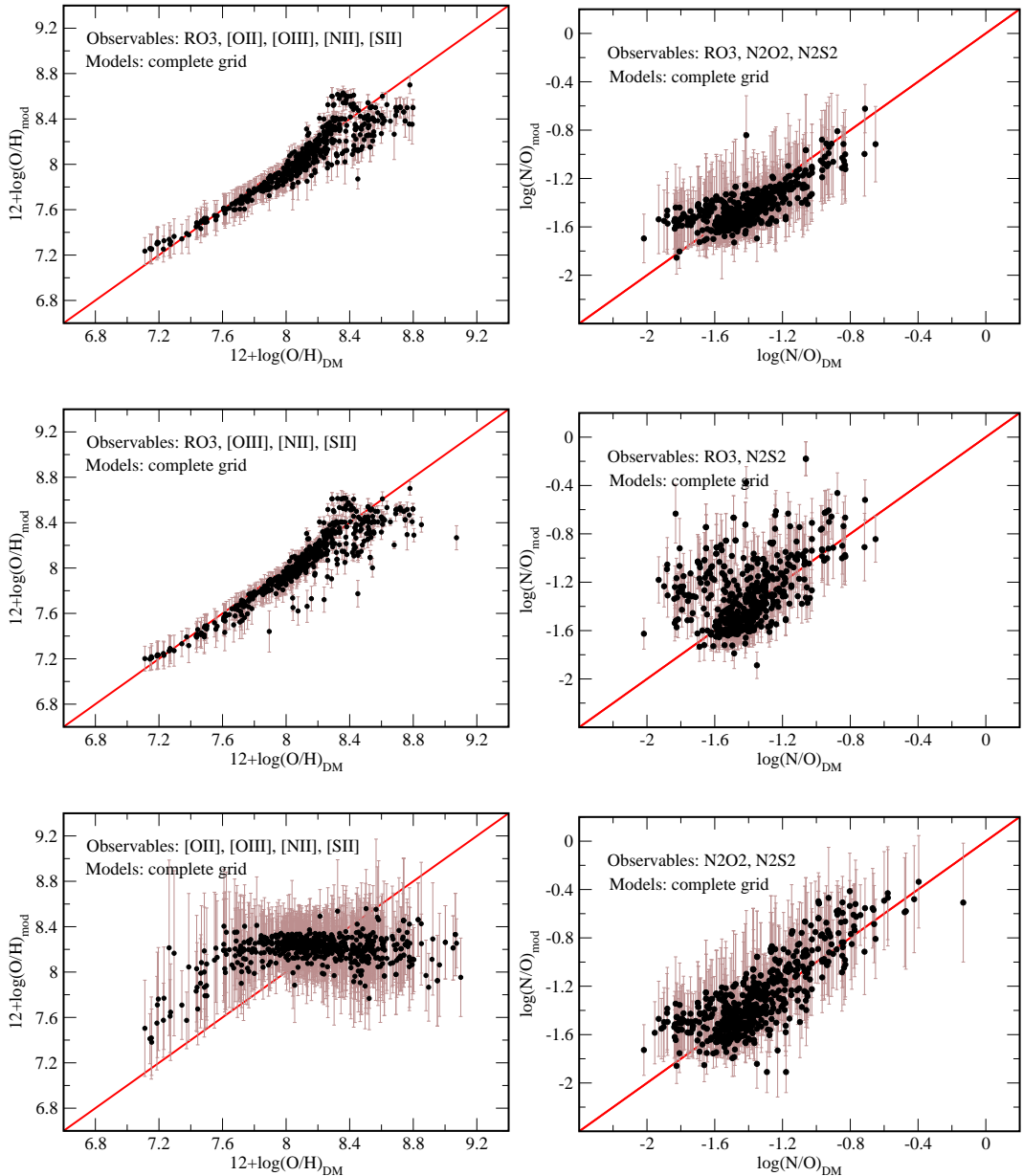

\begin{minipage}{180mm}
\centerline{
\psfig{figure=OH_comp_4363_all-grid.eps,width=7cm,clip=}
\psfig{figure=NO_comp_4363_all-grid.eps,width=7cm,clip=}}
\vspace{0.5cm}
\centerline{
\psfig{figure=OH_comp_sinOII_all-grid.eps,width=7cm,clip=}
\psfig{figure=NO_comp_sinOII_all-grid.eps,width=7cm,clip=}}
\vspace{0.5cm}
\centerline{
\psfig{figure=OH_comp_sin4363_all-grid.eps,width=7cm,clip=}
\psfig{figure=NO_comp_sin4363_all-grid.eps,width=7cm,clip=}}

\caption[]{Comparison between the oxygen abundances (left column) and
N/O ratio (right column) for the objects compiled as calculated following the
direct method described in the text and using the complete grid of models.
In the first row all lines ([\oii], [\oiii] 4363, 5007, [\nii], and [\sii]) are used. In the
middle, all lines except [\oii] and in the bottom row, all lines except [\oiii] 4363. 
The red solid line indicates in all plots the 1:1 relation.}
\label{OH_comp1}

\end{minipage}
\end{figure*} 

\begin{figure*}
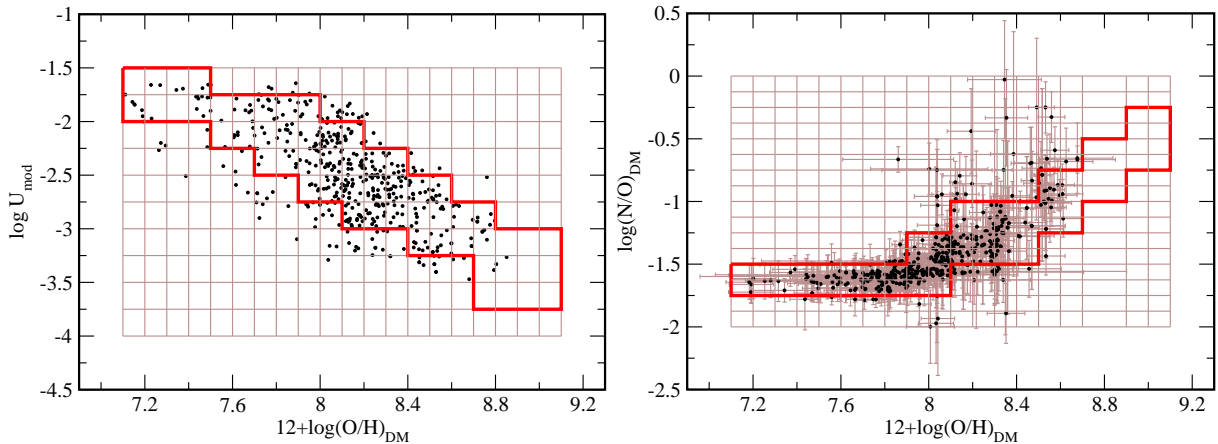

\begin{minipage}{180mm}
\centerline{
\psfig{figure=OH_logU_4363.eps,width=8cm,clip=}
\psfig{figure=OH_NO_4363.eps,width=8cm,clip=}}

\caption[]{At left, relation between 12+log(O/H) and log $U$ for the sample
of studied objects as derived using model-based values with the complete grid.
At right it is shown the relation with log(N/O). The space inside the red solid line 
in both panels indicates 
the limits of the empirically limited space of models described in the text.}
\label{OH_logU}

\end{minipage}
\end{figure*}

\section{Results and discussion}

\subsection{Abundances using the complete grid of models}

The procedure described in the previous section was followed to
derive O/H, N/O, and log $U$ from the models. The derived 
abundances were later compared with those  
calculated using the direct method in the sample of objects
compiled by \cite{RAM13}. The model-based abundances
were estimated using the five available emission lines and the complete
grid of models covering all possible excitation conditions. This excludes
the objects with higher $Z$ in the sample, as their abundances were derived
using other auroral lines than [\oiii] 4363 \AA.

In the upper panels of Figure \ref{OH_comp1} are shown the comparisons between
O/H (at left) and N/O (at right) as derived using this method and the
values obtained from the direct method as described in section 2.
In the case of O/H, as can be seen, the agreement is very good. The average of the
residuals to the 1:1 relation is lower than 0.1dex.
The dispersion, estimated as the standard deviation of the residuals, 
is lower for low-$Z$ (0.07dex for 
12+log(O/H)$<$8.0), even lower than 
the typical uncertainty in the derivation of the abundances following the
direct method ($\sim$ 0.1dex). 
Howevr, this dispersion is higher for high-$Z$ (0.14dex at
12+log(O/H)$>$8.0), mainly because the determination of O/H depends strongly
on the [\oiii] lines, which are much brighter in the low-$Z$ regime.
The dispersion in the case of N/O is of 0.15dex, and the models give
slightly higher values for very low N/O values. This is likely due to 
deviations of the assumption made to derive this ratio (N/O $\approx$
N$^+$/O$^+$), which is not always valid and depends on low-excitation lines
in a regime occupied mainly by low-$Z$ objects.

It is important to remark that no additional efforts were made to obtain this
agreement between the abundances derived from the models and the abundances derived
from the direct method, but it arises in a natural way when a consistent set of atomic data,
realistic geometrical conditions and an updated code and SEDs are used.

In the middle panels of the same figure, it is shown the comparison plots 
of abundances when
[\oii] 3727 \AA\ is not considered in the model-based abundances.
This situation happens in some spectral coverage configurations, such
as, for instance, the Sloan Digital Sky Survey (SDSS) spectra
for objects at a redshift lower than 0.02. As can be seen, for O/H
the agreement is still very good and, in fact, the dispersions both for
low-$Z$ (0.06dex) and high-$Z$ (0.12dex) are even better, always with
an average value of the residuals better than 0.1dex. This is mainly due to
the average O/H is not strongly affected by the relative emission of [\oii]
when all the [\oiii], both auroral and nebular, are available. The situation, however,
is sensibly worst in the case of N/O, where a dispersion of 0.25dex is found, with a high
dispersion for low N/O values. Since the main parameter to  
derive N/O is the N2S2 parameter, when [\oii] is not available, 
the contamination of the [\sii] emission lines in
low-$Z$ objects make this determination very uncertain. Anyway, this dispersion
is even better than the empirical dispersion of the N2S2 parameter found by \cite{PMC09}.

Since the main aim of this work is to provide alternative methodologies to
derive abundances consistent with the direct method when no auroral line
can be measured, the model-based abundances were calculated without
considering the [\oiii] 4363 \AA\ emission line. Notice that in this case, 
all the objects compiled by \cite{RAM13} with a direct determination
of the abundance are now considered in the analysis, even those of very high-$Z$ with other
auroral lines.
The results of this exercise are shown both for O/H and N/O in the
bottom panels of Figure \ref{OH_comp1}. In the case of O/H this procedure is 
clearly not valid as most of the objects give a very similar value 
around 12+log(O/H) = 8.2 
regardless of their real metallicity. Only in the case of extremely metal
poor objects (XMP; 12+log(O/H) $<$ 7.65) a different trend appears.
This behaviour demonstrates two things: i) [\oiii] 4363 \AA\ is the main 
discriminator between low-$Z$ and high-$Z$ objects independently 
of their excitation or geometrical conditions, and ii) very different conditions
of metallicity and/or excitation lead to similar sets of emission-line fluxes if
we consider with the same probability the whole space of possible values for
each parameter.
In the case of N/O, however, as the [\oii] emission line is now used, the agreement
improves and a dispersion of 0.22dex is found, about 0.1dex better
than any of the empirical calibrations of N2O2 or N2S2 found by \cite{PMC09}.

\begin{figure*}
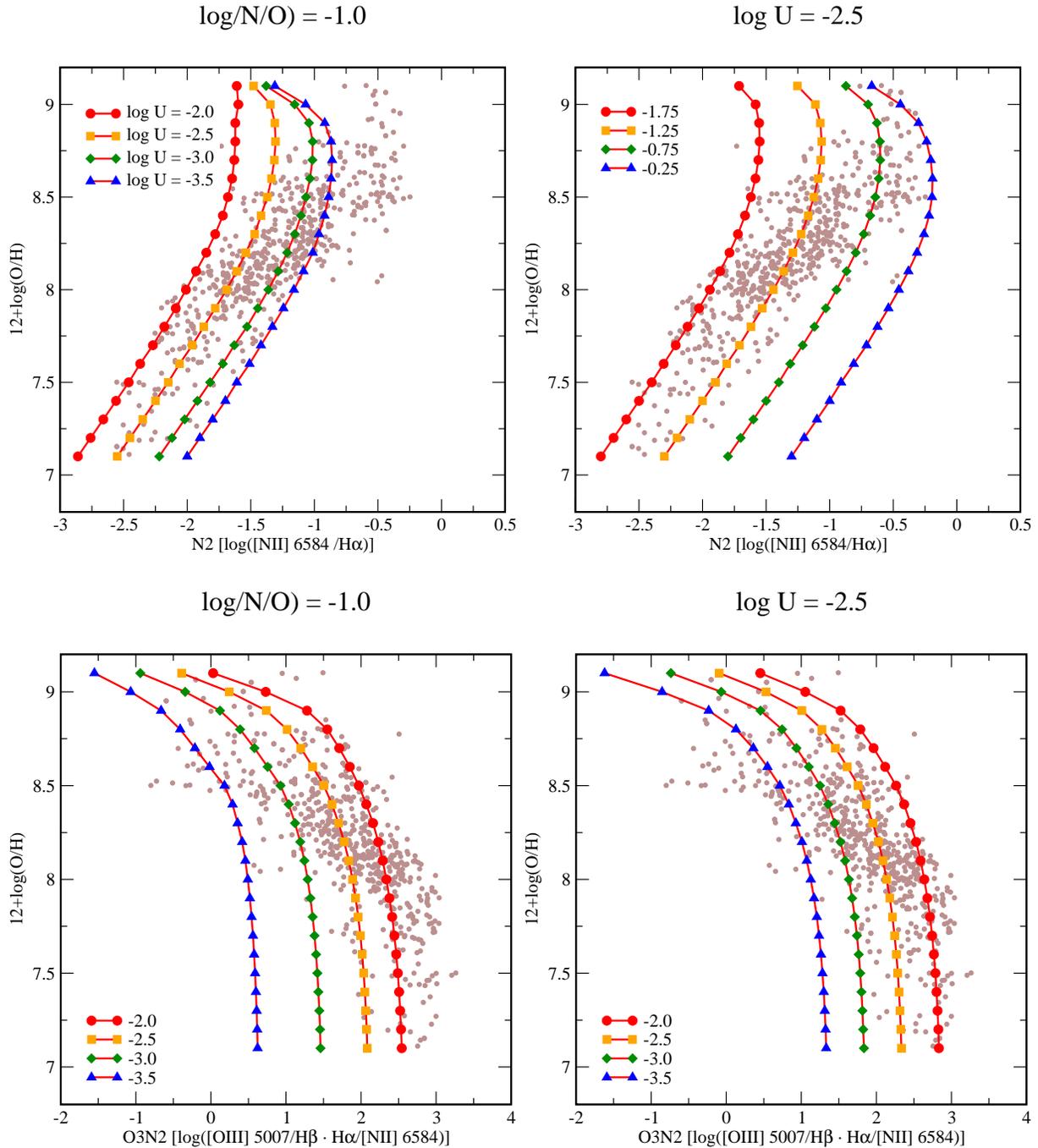

\begin{minipage}{180mm}
\centerline{
\psfig{figure=N2_OH_N1.0.eps,width=8cm,clip=}
\psfig{figure=N2_OH_U2.5.eps,width=8cm,clip=}}
\vspace{0.5cm}
\centerline{
\psfig{figure=O3N2_OH_N1.0.eps,width=8cm,clip=}
\psfig{figure=O3N2_OH_U2.5.eps,width=8cm,clip=}}

\caption[]{Relation between O/H and the N2 parameter in the upper row and with the O3N2 parameter
in the lower row for the sample of objects described in this work. Plots at the left column also show the
results from models for different values of log $U$ and a fixed log(N/O) = -1.0. In the right column,
the models have different values of N/O at a fixed log $U$ = -2.5.}
\label{OH_N2}

\end{minipage}
\end{figure*} 

\begin{figure*}
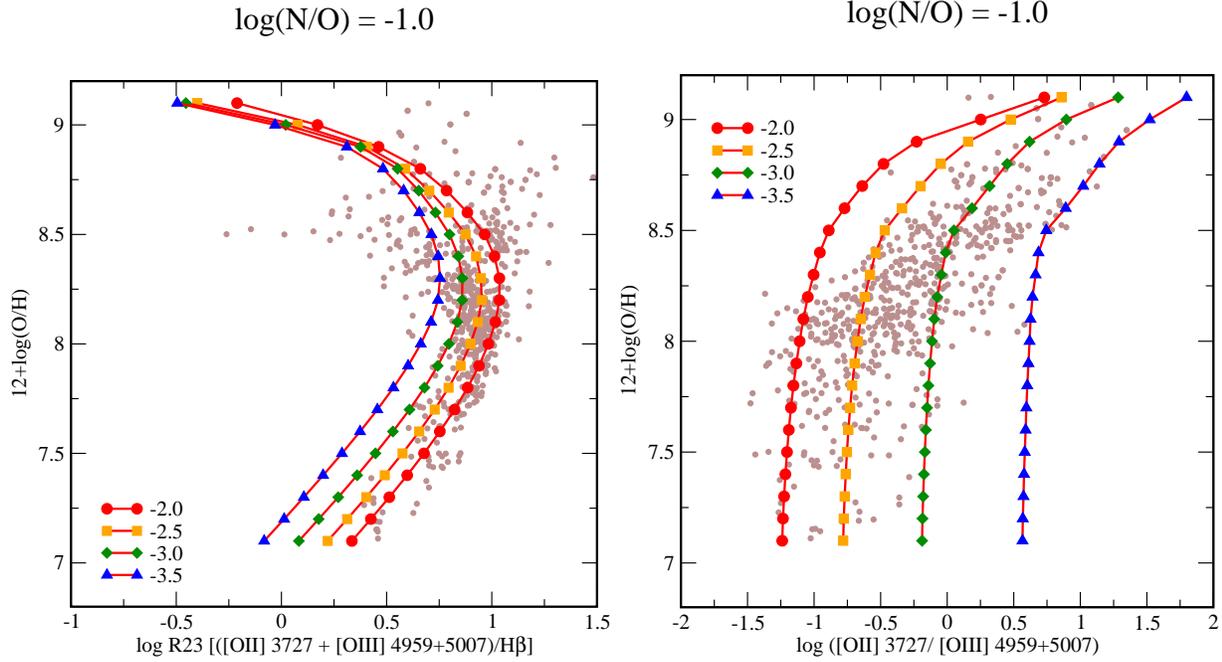

\begin{minipage}{180mm}
\centerline{
\psfig{figure=R23_OH_N1.0.eps,width=8cm,clip=}
\psfig{figure=O2O3_OH_N1.0.eps,width=8cm,clip=}}

\caption[]{Relation between O/H and the R23 parameter at left and the O2O3 parameter
at right for the sample of objects described in this work. Plots show the
results from models for different values of log $U$ and a fixed log(N/O) = -1.0.}
\label{OH_R23}

\end{minipage}
\end{figure*} 

\begin{figure*}
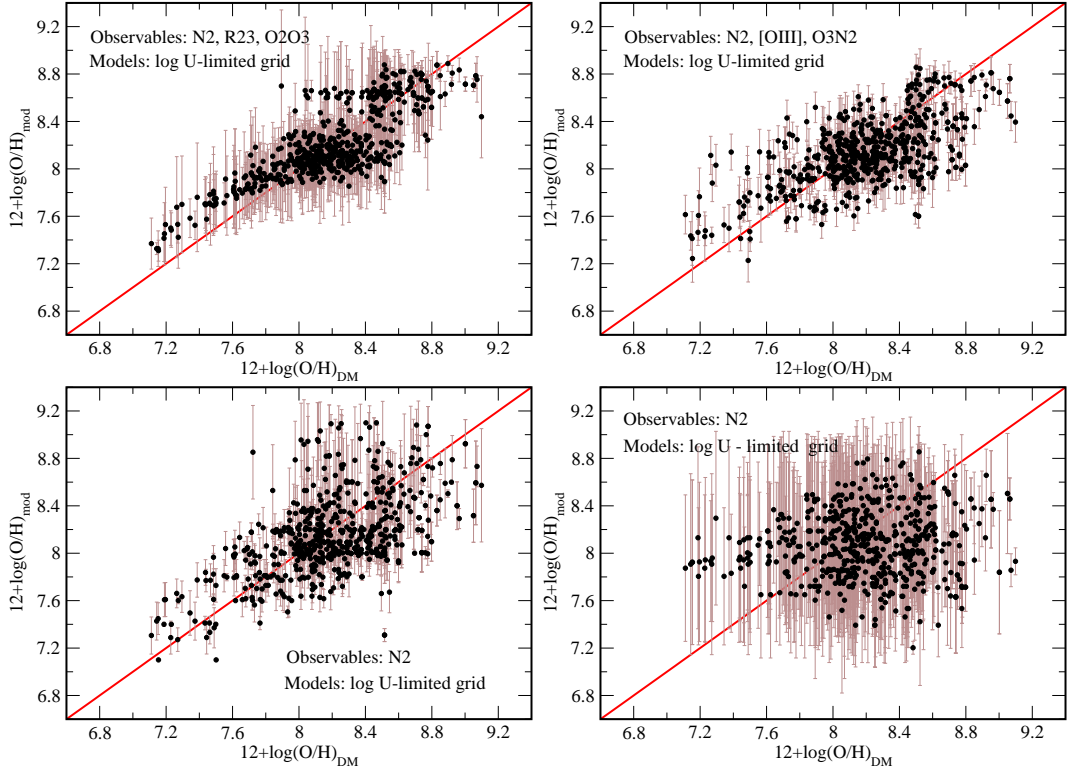

\begin{minipage}{180mm}
\centerline{
\psfig{figure=OH_comp_sin4363_logU-grid.eps,width=7cm,clip=}
\psfig{figure=OH_comp_sin4363-OII_logU-grid.eps,width=7cm,clip=}}
\centerline{
\psfig{figure=OH_comp_sin4363-OII-OIII_logU-grid.eps,width=7cm,clip=}
\psfig{figure=OH_comp_NII_logU-grid.eps,width=7cm,clip=}}

\caption[]{Comparison between the O/H derived from
the direct method and the model-based values using a grid with a log $U$ empirically 
limited. The upper left panel uses N2, R23, and O2O3. The upper right plot uses 
N2, [\oiii]/\hb\ and O3N2. The left lower panel only uses N2, but with a previous
limitation of N/O using N2S2. Finally, the right lower panel shows the comparison using
only N2 with no N/O restriction.}

\label{OH_comp2}

\end{minipage}
\end{figure*} 

\begin{figure*}
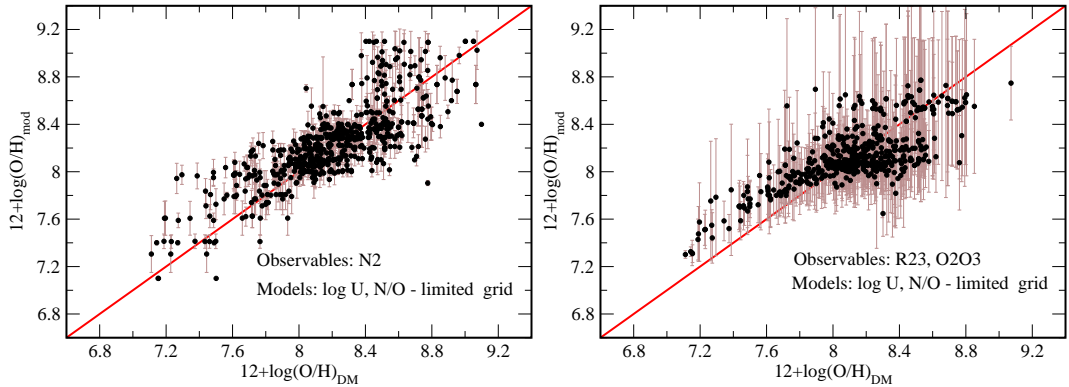

\begin{minipage}{180mm}
\centerline{
\psfig{figure=OH_comp_NII_logU-NO-grid.eps,width=7cm,clip=}
\psfig{figure=OH_comp_OII-OIII_logU-NO-grid.eps,width=7cm,clip=}}

\caption[]{Comparison between O/H derived from the direct method and the
model-based values using a grid empirically limited both for log $U$ and log (N/O).
At left, using the N2 parameter, while at right using R23 and O2O3.}

\label{OH_comp3}
\end{minipage}
\end{figure*}

\subsection{Abundances using a log $U$ empirically limited grid of models}

In order to give reliable model-based abundances consistent with the
direct method in absence of the [\oiii] 4363 \AA\ auroral line, and 
therefore let this method to work in high-$Z$, faint or high redshift
star-forming objects two important assumptions are made.
Firstly, the space of possible excitation conditions and metallicity is
restricted empirically to fit the trend obtained for the studied sample,
for which reliable values of log $U$ are obtained when using all the
emission lines. This empirical relation between O/H and log $U$ is
plotted in left panel of Figure \ref{OH_logU} along with a solid line
that encompasses the space of considered combinations. As can be seen,
there is a trend to find higher values of log $U$ at lower metallicities,
while the opposite is true at high-$Z$ (the coefficient of correlation is -0.63). 
Although there are objects that
lie outside the assumed possible values, the majority of them
lie in a region that minimizes the dispersion in the final derivation of
O/H.

The second approximation that improves the agreement between the
model-based O/H abundances and those derived from the direct method is to
change the set of observables considered to calculate $\chi^2$ in
equation \ref{chi}. In this case, emission-line ratios known to have a
clear dependence on $Z$ or log $U$ are used. This is the case
of [\nii]/\hb\ (or equivalently [\nii]/\ha) defined as the N2 parameter \citep{storchi,D02},
which has a known dependence on $Z$ as can
be seen in upper panels of Figure \ref{OH_N2}. 
The different grids of models to show the high dependence of this parameter 
on log $U$ and N/O as already pointed out by \cite{PMD05}. Notice as well that
although models predict that N2 is insensitive to O/H up to values
12+log(O/H) $\approx$ 8.5, the empirical calibrations of this parameter such
as in \cite{PP04} or \cite{PMC09} can work up to values twice the
solar value (around 12+log(O/H) = 9.0) because of the empirical relation
between $Z$ and log $U$ or with N/O.

In the case of [\oii] and [\oiii] lines the model-based abundances better agree with
those from the direct method if combinations of these two lines are used as observables,
such as R23 \citep{Pagel79}, (([\oii]+[\oiii])/\hb), which has a bi-valuated 
behaviours in its dependence with $Z$ (see left panel of Figure \ref{OH_R23}) 
and also presents a high dependence on log $U$ and on the effective ionising
temperature \citep{PMD05}. This dependence can be partially reduced using
the O2O3 parameter, defined as [\oii]/[\oiii] as used by \cite{kobul99} in their
fittings of the \cite{McG91} models or by \cite{Pilyugin05} using a different
formalism with their $P$ parameter. In right panel of Figure \ref{OH_R23} it
is shown the slight dependence of this ratio on $Z$ so it can be used to reduce
the dependence of R23 on it.

In this way, using these three observables and the grid of models
for those combinations of $Z$ and log $U$ limited by the studied sample,
a much better agreement between the model-based abundances
and those obtained from the direct method is obtained, as can be seen in upper left
panel of Figure \ref{OH_comp2}. Although the agreement is now much better, the situation is
different in each metallicity regime. While for low-$Z$
(12+log(O/H) $<$ 8.0), the dispersion is lower (0.16dex), there
is a systematic offset of about 0.2dex to find larger metallicities using
the models. The disagreement at this regime can be
possibly due to the low-excitation emission from the diffuse
gas or perhaps low-velocity shocks that increase the expected
flux of these lines.
However, this is not critical taking into account that i) according to
\cite{PM13} less than 1 per cent of star-forming galaxies lie in this regime and
ii) the [\oiii] 4363 {\AA} is prominent and easy to measure at these metallicities.
On the contrary, for 12+log(O/H) $>$ 8.0, the agreement between the
metallicities of the direct method and the model-based values is better than
0.05dex, the usual uncertainty associated with the abundances, but
the dispersion enhances up to 0.19dex.

Considering again the case when [\oii] 3727 {\AA} is not available
(e.g. SDSS spectra of low-$z$ objects), the observables must be
redefined. A good alternative to O2O3 is the O3N2 parameter, defined
as the ratio of [\oiii]/\hb\ and [\nii]/\ha\ and used as a
estimator of metallicity originally by \cite{alloin79}. In the lower panels
of Figure \ref{OH_N2} it is shown the dependence of this parameter with oxygen
abundance using the sample of compiled objects and exploring its behaviour for
fixed values of log $U$ and log(N/O). The dependence of this
parameter on N/O is reduced in the models for each object with the
previous estimation of this ratio using the N2S2 parameter. Hence in
this case for the estimation of oxygen abundance, assuming only certain values of
log $U$ the observables are [\nii], [\oiii], and O3N2. The metallicities obtained using
this procedure are shown in the upper right panel of Figure \ref{OH_comp2}.
As in the case with [\oii], the agreement is better for high-$Z$ than for
low-$Z$, where model-based O/H are in average 0.29dex larger. Besides, the
dispersion is very similar in the two regimes (0.21dex and 0.24dex respectively), 
because O3N2 tends to overestimate O/H in the low-$Z$ regime.

In lower left panel of Figure \ref{OH_comp2} it is shown the comparison between the
abundances from the direct method and those from the models when only [\nii] and [\sii]
emission lines are used. In this case, N/O is firstly estimated using N2S2 and later, once the 
grid limited to the closest values of N/O and the empirical values of log $U$,
O/H is also derived by means of only [\nii]/\hb. The dispersion of this
comparison is slightly better than the value obtained in the empirical calibration of the
N2 parameter (0.31dex, \cite{PMC09}), but as in the previous cases, the
model-based O/H values in the low-$Z$ regime are systematically higher, with
an average difference of 0.4dex, what supports the idea that low-excitation lines are 
overestimated in this regime.

Finally, in the right lower panel of Figure \ref{OH_comp2} it is shown the comparison
using only [\nii]/\hb, when no previous estimate of N/O is done. In this case,
no linear correlation at all is found, as all possible values of N/O are considered
in the weighted average of O/H. Therefore, no reliable estimation of O/H 
can be done using only [\nii] in relation to a Balmer hydrogen emission line if
no previous guess about the value of N/O is done.

\subsection{Abundances using a N/O empirically limited grid of models}

In order to improve the agreement between the oxygen abundance derived using
the direct method and the model-based values when only [\nii] emission 
line is available, what is the case for many high redshift star-forming objects observed
in the IR, it is assumed that only certain values for N/O are valid in each
metallicity regime. Of course, the use of this limited grid implies the implicit
and not necessarily correct assumption that the studied object has in average
the same properties that the sample studied here. However, there are cases
in the literature, where combinations of N/O and O/H do not follow or lie out 
of the trends shown by the most part of star-forming regions/galaxies. This is the case, for instance, of {\em green
pea} galaxies \citep{Amorin10,Amorin12}, which present very low-$Z$ values 
and almost solar N/O. This, according to the authors, could be indicative that
the massive star-formation processes taking place in these objects can be due
to inflows of pristine gas and/or outflows of enriched material. The 
importance of these mechanisms should not be neglected as these are thought
to be behind the empirical law found between metallicity and star formation rate
(e.g. \citealt{lara10,mannucci10,PM13}).
Therefore, it is convenient to have a reliable estimation of N/O before deriving O/H only
with the aid of [\nii] emission lines.

In the right panel of Figure \ref{OH_logU} it is shown the relation between O/H and N/O
derived using the direct method in the compiled sample of objects. The solid line encompasses the
set of models considered to find abundances. As can be seen, very low values of log(N/O)
at a constant value of -1.75-1.50 are obtained at 12+log(O/H) $<$ 8.0, consistently with
the predictions of a production of primary N in this regime, while for high-$Z$ there
is a dependence of N/O with O/H, as the main production of N has a secondary origin
(e.g. \citealt{henry2000}). Since no data populate the high-$Z$ regime
of this diagram, it has been assumed that the linear relation between O/H and N/O
extends in this regime.

Using this new limitation in the grid of models, added to that already considered between
O/H and log $U$, new O/H values based only on [\nii] are derived and a better agreement is found but only for high-$Z$ with
a dispersion of 0.24dex. At low-$Z$, however, a systematic offset of more than
0.4dex is found with a huge dispersion.

In the right panel of Figure \ref{OH_comp3} it is shown the comparison when
only [\oii] and [\oiii] strong emission lines are used, which are almost insensitive
to N/O, so the N/O limited grid is more appropriate to derive reliable uncertainties.
Taking again as observables R23 and O2O3 and with a previous limitation of
the grid to the most probable values of log $U$, In this case,
the dispersion is of only 0.19dex at all $Z$, but with an overestimation
of O/H for the model-based values at 12+log(O/H) $<$ 8.0 of 0.26dex.

\subsection{Application to the abundance gradient in M101}

\begin{figure*}
\begin{minipage}{180mm}
\centerline{
\psfig{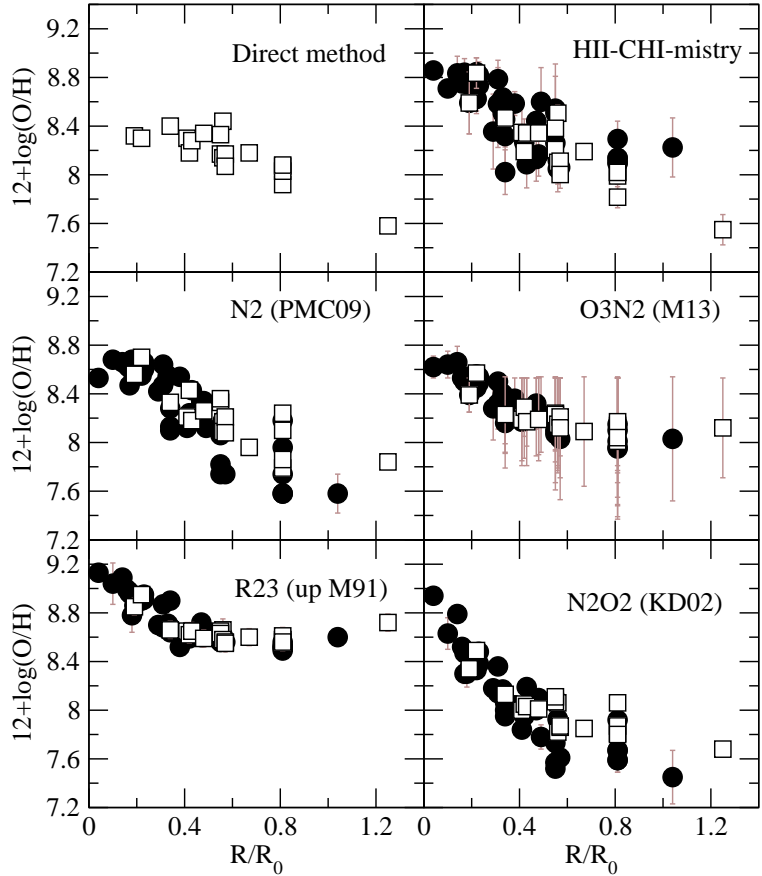}}

\caption[]{Galactocentric distribution of 12+log(O/H) as a function of the effective radius for
the \hii\ regions in M101 observed by \cite{k96} (black circles) and
\cite{k03} (white squares). The panels show the oxygen abundances as
derived using the following methods from left to right and from top to
bottom: the direct method, the model-based $\chi^2$ described in this
work, the N2 parameter calibrated by \cite{PMC09}, the O3N2 parameter
calibrated by \cite{RAM13}, the \rdostres\ calibrated by \cite{McG91}, and
the N2O2 parameter calibrated by \cite{KD02}.}

\label{m101_OH}

\end{minipage}
\end{figure*} 

\begin{figure*}
\begin{minipage}{180mm}
\centerline{
\psfig{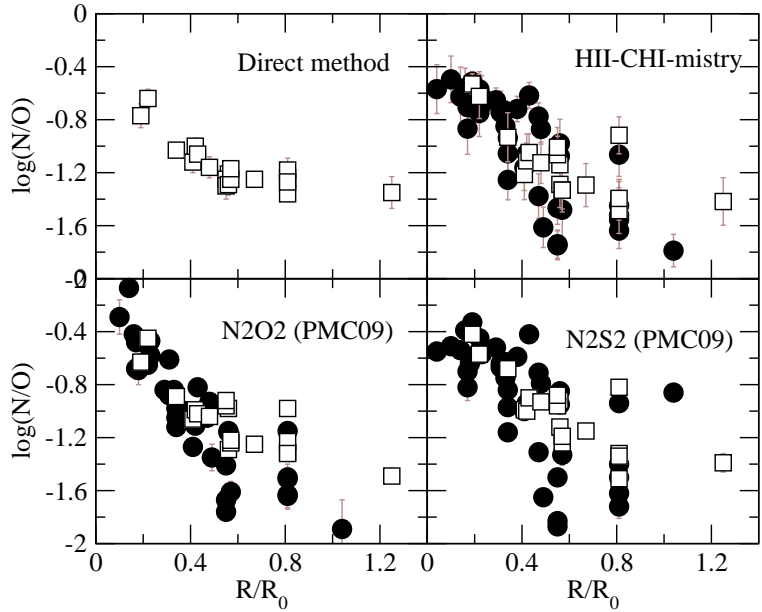}}

\caption[]{Galactocentric distribution of log(N/O) as a function of the effective radius for
the \hii\ regions in M101 observed by \cite{k96} (black circles) and
\cite{k03} (white squares). The panels show N/O ratios as
derived using the following methods from left to right and from up to
down: the direct method, the model-based $\chi^2$ described in this
work, and the N2O2 and N2S2 parameters calibrated by \cite{PMC09}.}

\label{m101_NO}

\end{minipage}
\end{figure*} 

The study of gradients of metallicity across spiral galaxies is one of the
issues where robust methods of determination of metallicity are needed, because
its variation can involve in the same galaxy high- and low-$Z$ \hii\ regions.
For instance, M101 is an object with a very prominent abundance gradient, 
with a variation of  more than an order of magnitude in oxygen abundance
from the inner to the outer regions. In order to evaluate the model-based
$\chi^2$ method described in this work, emission-line data 
from the \hii\ regions observed by \cite{k96} and \cite{k03} were compiled.
Later their O/H and N/O were obtained using different strong-line methods. 
In the case of 19 \hii\ regions described in \cite{k03} the direct method can also be
applied, as these authors provide good signal-to-noise measures of different
auroral lines. In this case, the expressions described in section 2 were used
to re-calculate electron temperatures and ionic abundances for this sample.

In Figure \ref{m101_OH} are shown the derived gradients in this galaxy
using different methods. All of them show a prominent O/H gradient, but the
behaviour can be very different depending on the used method. Using the $\chi^2$
model-based abundances, the maximum value is reached in the innermost regions, then
the gradient is flattened for R$>$0.8$\cdot$R$_0$ (NGC 5471), but the
oxygen abundance in the outermost position (SDH323, R/R$_0$ = 1.25)
is sensibly lower. The oxygen abundances derived from the direct method 
for those \hii\ regions with at least one auroral line is totally consistent
with this scenario, but not completely owing to the lack of points 
in the innermost regions and between NGC5471 and SDH323. In 
Figure \ref{m101_NO} are shown the gradients for the 
same \hii\ regions for N/O. Again, the agreement between the direct method and
the model-based abundances is very good, but the behaviour is sensibly
different to that of O/H in the outer positions, because there is not 
apparent flattening and the N/O ratio for SDH323 is sensibly higher than
expected. This very high N/O ratio could be consistent with an inflow of pristine
gas, making the metallicity to decrease but keeping at a higher level N/O.
This overabundance of N/O makes all gradients derived using [\nii] emission lines
(e.g. N2, O3N2, N2O2) to very high values of $Z$ in this \hii\ region. 
In fact, the $Z$ gradient derived from N2O2 is totally similar to that of N/O.
Besides, these parameters
lead to strange behaviours (e.g. lower $Z$ values in the innermost regions 
for N2 and O3N2 and also higher $Z$ in the outer regions for O3N2)
in those regimes of metallicity where they are not well calibrated.
In the case of \rdostres\ , as it was calibrated using not empirically constrained
models, the absolute value of oxygen is about 0.4dex higher than in the direct
method. Besides, if [\nii] is used to select the upper branch calibration, the
value for O/H can be even larger than in some inner positions. Finally, the N/O
gradient derived from other strong-line methods, such as N2O2 or N2S2, is
not very different from the models described here, but with a higher dispersion.

\section{Summary and conclusions}

This work analyses whether the predicted
intensities of the strongest collisionally excited optical lines emitted
by a gas ionised by massive stars
made by photoionisation models can yield chemical abundances consistent
with the direct method. For this purpose profiting the
compilation made by \cite{RAM13} of \hii\ regions and star-for
galaxies with auroral emission-lines, their oxygen and nitrogen abundances
were recalculated using the software {\sc pyneb} \citep{pyneb}. 
Besides, a large grid of {\sc cloudy} \citep{cloudy} photoionisation models was
computed and their relative optical emission lines, O/H, N/O, and log $U$
were collected. Using a $\chi^2$ weighted mean procedure the
chemical abundances for each observed object was recalculated using the
models.

In a first step a model-based N/O is found using emission-line indicators
sensitive only to this abundance ratio (i.e. N2O2, N2S2). 
Once N/O is enclosed, O/H and log $U$ are searched in a new
iteration. This procedure, publicly available through a
script called {\sc HII-CHI-mistry}, leads to the following conclusions:

- The agreement between the model-based O/H and N/O abundances and those
derived using the direct method is excellent when all explored lines
are used ([\oii] 3727 \AA, [\oiii] 4363, 5007 \AA\AA, [\nii] 6584 \AA, and [\sii] 6717+6731 \AA\
all relative to \hb).  This agreement arises in a natural way and no additional
efforts were made to force this match.
The dispersion for O/H is slightly higher
for 12+log(O/H) $>$ 8.0 probably owing to that the determination depends
mostly on [\oiii] emission lines. The agreement is quite similar even when
[\oii] emission line are not considered. 

- When using the grid of models covering all possible combinations of O/H, N/O,
and log $U$ no reliable model-based estimation of O/H can be obtained if
[\oiii] 4363 \AA\ is not taken into account. However, assuming an empirical
relation between O/H and log $U$ and considering $Z$-sensitive observables
in the $\chi^2$ method a very good agreement is obtained for 12+log(O/H) $>$ 8.0,
where more than 99\% of star-forming objects lie and where it is more difficult
to detect the {\oiii] 4363\AA).
At low-$Z$ there is a systematic offset to obtain higher
values of O/H according to models, possibly related with the 
contamination of the low excitation emission lines (diffuse gas and/or low velocity
shocks).

- The dispersion in the comparison between model-based abundances and
those obtained from the direct method is worse as a lower number of emission-lines
is considered, but the obtained are always better than in other empirical
calibrations using the same involved lines.
When no previous estimation of N/O can be made by means
of N2O2 or N2S2, additional assumptions about the O/H vs. N/O,
not necessarily always true, should be made
in order to obtain O/H values only from [\nii]/\ha. 

- The recipes to enclose the grid of models to derive metallicities when no auroral
lines are available are arbitrary and based on an empirical set of data of the local Universe.
This problem is similar to the that pointed out by \cite{stasinska10} for the
empirical calibration of certain strong-line methods.
Other recipes can be applied for the enclosing of the three input parameters (O/H, N/O, log $U$) 
if a limited set of lines is available and  possible different scenarios are envisaged in order to
arrive to realistic derivations of the chemical properties of the studied objects.
Alternative recipes can be applied to the used SEDs and cluster ages used in this work.

- The use of this procedure in objects with no detection of any auroral
line (faint objects, high redshift star-forming galaxies, high-$Z$ \hii\ regions) 
to derive chemical abundances can lead to values consistent with the
direct method instead of strong-line methods. The use of these strong-line
methods has the disadvantage that they are usually calibrated using a limited sample
of objects or grids of models that do not cover all possible physical conditions.
This implies that later comparisons with objects whose abundances were calculated using
other methods are often inconsistent and present non-negligible offsets between
them.

- The new method based on empirically constrained models was applied to
study the gradients of O/H and N/O in M101. The results are totally consistent with
the direct method in those regions with at least one auroral line and it is robust both
in the innermost high-$Z$ and the outer low-$Z$ positions, where other strong-line
methods are not well calibrated or suffer very high dependence on other ionic ratios,
such as N/O.

\section*{Acknowledgements}
The author would like to thank Raffaella A. Marino on behalf of the CALIFA collaboration
for giving public access to the database of emission-lines in objects with measurements of 
the auroral lines.
This work has been partially supported by research project AYA2010-21887-C04-01
of the Spanish National Plan for Astronomy and Astrophysics, and by the projects 
PEX2011-FQM7058 and TIC114  {\em Galaxias y Cosmolog\'\i a} of the
 Junta de Andaluc\'\i a (Spain).
The author also acknowledges Jos\'e M. V\'\i lchez, Rub\'en Garc\'\i a-Benito,
Ricardo O. Amor\'in, Christophe Morisset, Thierry Contini, and Xuan Fang for many fruitful discussions
and suggestions that have helped to conceive and improve this manuscript. 
The manuscript also benefited from the constructive suggestions and
comments from an anonymous referee.

\bibliographystyle{mn2e}
\bibliography{HII-CHI-mistry}

\end{document}